\documentclass[a4paper,fleqn]{cas-dc}

\usepackage[authoryear]{natbib}

\usepackage{amsmath,amssymb,amsfonts}
\usepackage{algorithmic}
\usepackage{graphicx}
\usepackage{textcomp}
\usepackage{ragged2e}
\usepackage{balance}
\usepackage{colortbl}
\usepackage{multirow}
\usepackage{bm}
\usepackage{balance}
\usepackage[figuresright]{rotating}
\usepackage{rotfloat}
\usepackage{float}
\usepackage{enumerate}
\usepackage{framed}
\usepackage{xcolor}
\usepackage{booktabs}
\usepackage{caption}
\usepackage{tabularx} 
\usepackage{multirow}
\usepackage{makecell}
\usepackage{stfloats}
\usepackage{subfigure}
\usepackage{multirow, hhline}
\usepackage{CJKutf8}

\definecolor{backcolor}{rgb}{0.95,0.95,0.95}

\usepackage{listings}
\usepackage{xcolor}

\def\tsc#1{\csdef{#1}{\textsc{\lowercase{#1}}\xspace}}
\tsc{WGM}
\tsc{QE}
\tsc{EP}
\tsc{PMS}
\tsc{BEC}
\tsc{DE}

\begin{document}
\begin{sloppypar}

\let\WriteBookmarks\relax
\def\floatpagepagefraction{1}
\def\textpagefraction{.001}

\shorttitle{Unveiling Security Weaknesses in Autonomous Driving Systems}
\shortauthors{Z Li et al.}
\title [mode = title]{Unveiling Security Weaknesses in Autonomous Driving Systems: An In-Depth Empirical Study}

\author[1]{Wenyuan Cheng}
\ead{closerecover@mails.ccnu.edu.cn}  

\credit{Conceptualization, Methodology, Investigation, Data curation, Writing - Original draft preparation}

\address[1]{School of Computer Science \& Hubei Provincial Key Laboratory of Artificial Intelligence and Smart Learning, \\Central China Normal University, Wuhan, China \\}

\author[1]{Zengyang Li}
\cormark[1]
\ead{zengyangli@ccnu.edu.cn}
\credit{Conceptualization, Methodology, Investigation, Data curation, Writing - Original draft preparation}

\author[2]{Peng Liang}
\ead{liangp@whu.edu.cn}
\credit{Conceptualization, Methodology, Writing - Original draft preparation}
\address[2]{School of Computer Science, Wuhan University, Wuhan, China}

\author[1]{Ran Mo}
\ead{moran@ccnu.edu.cn}
\credit{Conceptualization, Methodology, Writing - Original draft preparation}

\author[3]{Hui Liu}
\ead{hliu@hust.edu.cn}
\credit{Conceptualization, Methodology, Writing - Original draft preparation}
\address[3]{School of Artificial Intelligence and Automation, Huazhong University of Science and Technology, Wuhan, China}

\cortext[cor1]{Corresponding author.}

\begin{abstract}
\noindent \textbf{Context}:
The advent of Autonomous Driving Systems (ADS) has marked a significant shift towards intelligent transportation, with implications for public safety and traffic efficiency. While these systems integrate a variety of technologies and offer numerous benefits, their security is paramount, as vulnerabilities can have severe consequences for safety and trust. \\ 
\noindent\textbf{Objective}: This study aims to systematically investigate potential security weaknesses in the codebases of prominent open-source ADS projects using CodeQL, a static code analysis tool. The goal is to identify common vulnerabilities, their distribution and persistence across versions to enhance the security of ADS. \\ 
\noindent\textbf{Methods}: We selected three representative open-source ADS projects, Autoware, AirSim, and Apollo, based on their high GitHub star counts and Level 4 autonomous driving capabilities. Using CodeQL, we analyzed multiple versions of these projects to identify vulnerabilities, focusing on CWE categories such as CWE-190 (Integer Overflow or Wraparound) and CWE-20 (Improper Input Validation). We also tracked the lifecycle of these vulnerabilities across software versions. This approach allows us to systematically analyze vulnerabilities in projects, which has not been extensively explored in previous ADS research. \\ 
\noindent\textbf{Results}: Our analysis revealed that specific CWE categories, particularly CWE-190 (59.6\%) and CWE-20 (16.1\%), were prevalent across the selected ADS projects. These vulnerabilities often persisted for over six months, spanning multiple version iterations. The empirical assessment showed a direct link between the severity of these vulnerabilities and their tangible effects on ADS performance. \\ 
\noindent\textbf{Conclusions}: These security issues among ADS still remain to be resolved. Our findings highlight the need for integrating static code analysis into ADS development to detect and mitigate common vulnerabilities. Meanwhile, proactive protection strategies, such as regular update of third-party libraries, are essential to improve ADS security. And regulatory bodies can play a crucial role in promoting the use of static code analysis tools and setting industry security standards. \\ 

\end{abstract}


\maketitle

\section{Introduction} \label{Introduction}
The rapid proliferation of autonomous driving technologies in recent years has revolutionized the transportation sector, marking a significant shift toward intelligent vehicle operation~\citep{yaqoob2019autonomous}. Autonomous vehicles (AVs) represent a complex integration of various hardware and software technologies, including perception, decision-making, and seamless interaction with cloud platforms for high-definition map generation and data storage~\citep{lee2020design}. Despite technological advancements, concerns regarding the safety of AVs have been raised, with some fears being manifested in recent accidents involving partially automated systems~\citep{casner2016challenges, patel2024exploratory}. These incidents underscore the imperative for adopting more diverse and efficient testing methodologies to rigorously evaluate and enhance the security of autonomous driving systems (ADS\footnote{ADS can be singular or plural depending on the context.}) before they become commonplace on our roads. Our study aims to address this gap by systematically investigating the code vulnerabilities in prominent open-source ADS projects using CodeQL, a static code analysis tool.

Code vulnerabilities are an inevitable aspect of software development, and developers inadvertently introduce myriad vulnerabilities into their codebases~\citep{iannone2022secret}. These vulnerabilities often go unnoticed due to the complexity of modern software systems and the limited visibility developers have in all aspects of their projects~\citep{li2022vulnerability}. Although not all vulnerabilities lead to system crashes, they can result in information leaks or tampering, which may be considered severe in their implications~\citep{votipka2020understanding}. In the context of ADS, the stakes are significantly higher, as these systems interact directly with individuals who may not possess technical knowledge to address security concerns. Consequently, a more rigorous security standard is required for ADS, necessitating a stringent review process during their development to mitigate the risks associated with potential vulnerabilities~\citep{liu2020computing}. This is exactly where static code analysis can play its role.

In the landscape of software development, the management and mitigation of vulnerabilities within open-source projects have become increasingly critical, and researchers have shed light on the fact that a significant portion of vulnerabilities could be avoided with adequate internal testing~\citep{bandara2021large, ayala2023empirical}. Static code analysis plays a pivotal role in this regard, as it allows the examination of code without execution, thereby identifying potential security flaws that could lead to vulnerabilities~\citep{ayala2023empirical}. This method stands in contrast to dynamic code analysis that requires the running of the code to detect vulnerabilities. Meanwhile, it offers the advantage of being able to catch issues that may not manifest during runtime~\citep{bandara2021large}. Moreover, static code analysis promotes a culture of writing secure code from the onset, establishing a quality and security baseline for a project between team members
. Despite these advantages, the adoption of static analysis in open-source projects, particularly those with collaborative environments, is not as widespread as one might expect. This calls for a more concerted effort to integrate static code analysis into the development workflows of such projects, ensuring a more secure and reliable software ecosystem.

CodeQL, as highlighted by~\cite{ayala2023empirical, szabo2023incrementalizing}, stands out among static code analysis tools due to its unique declarative approach, which allows the specification of rules and queries to identify not only security vulnerabilities but also code issues that affect the availability or stability of a system. This approach extracts facts from source code and evaluates them against a set of user-defined rules, providing a more targeted and comprehensive analysis compared to traditional static analysis tools. The studies above advocate for a broader adoption of CodeQL, suggesting that its advanced features and flexibility could greatly contribute to the mitigation of software vulnerabilities if utilized more extensively across various projects.

Given the intricacies and stringent security demands of ADS, it is imperative to adopt a comprehensive suite of technologies to protect against the introduction of vulnerabilities during development. This paper is dedicated to leveraging CodeQL to probe into and track potential code vulnerabilities within three prominent open-source ADS projects: Autoware, AirSim, and Apollo. These projects are primarily composed of code written in C++, Python, and JavaScript, which are most commonly used in ADS for high-performance computing (C++), AI-based planning (Python), and frontend construction (JavaScript)~\citep{bathla2022autonomous, lin2018architectural, tiobe_index, octoverse_2024}. C++ serves as the predominant programming language in the three projects, with Autoware (Autoware.universe) consisting of 92.1\% C++, Apollo 74.5\%, and AirSim 73.7\%~\citep{autowarefoundation_autoware, autowarefoundation_autoware_universe, apolloauto_apollo, microsoft_airsim}. This uniformity in programming languages ensures a consistent and comparable environment for analysis. We focuses on three key aspects: the patterns in which vulnerabilities are distributed across the selected ADS projects, the longevity of these vulnerabilities before they are addressed, and the extent of how the identified vulnerabilities affect the performance of ADS. Unlike previous studies that focused on commits rather than directly examining the codebases~\citep{garcia2020comprehensive}, we aim to identify common vulnerabilities, their distribution across modules, and their persistence across versions with more detailed and comprehensive analysis to improve the security of ADS.

Our study revealed that specific CWE categories, particularly CWE-190 (Integer Overflow or Wraparound, 59.6\% in total) and CWE-20 (Improper Input Validation, 16.1\% in total), were prevalent across the selected ADS projects. The vulnerabilities were concentrated in perception-related modules and often persisted for over six months, spanning multiple version iterations. The empirical assessment showed a direct link between the severity of these vulnerabilities and their tangible effects on ADS performance, which underscores the need for integrating static code analysis into ADS development workflows.

The \textbf{contributions} of this study are threefold. Firstly, it aims to enhance the systematic examination of ADS security, offering insights into the distribution and persistence of vulnerabilities in a domain where the consequences of failure are significant. Secondly, by employing CodeQL, this work aims to demonstrate the effectiveness and practicality of the tool in uncovering security weaknesses in large and complex software systems. Lastly, this study intends to provide developers and researchers with evidence-based recommendations for improving the security of ADS, drawing from the empirical findings of the analysis. Through these contributions, the paper aspires to advance the field of autonomous driving by integrating empirical research with actionable strategies, thereby bolstering the security and reliability of ADS. 

The rest of this paper is structured as follows: Section \ref{Research_Design} delves into research design and methodology of this study; Section \ref{Results} presents the study results; Section \ref{Discussion} analyzes the results and discusses their implications; Section \ref{Threats_to_validity} identifies threats to the validity of the results; Section \ref{related_work} outlines the related work; and Section \ref{Conclusions} concludes this work with future research directions.

\section{Research Design} \label{Research_Design}

\subsection{Research Questions}

The \textbf{objective} of this research is to \textit{analyze} potential security weaknesses in the codebases of prominent ADS projects using CodeQL, \textit{for the purpose of} identifying and analyzing common vulnerabilities across versions, \textit{with respect to} enhancing the security of ADS \textit{in the context of} open-source software development.

To guide this exploration, we have formulated a set of research questions (RQs) that focus on the distribution, life cycle, and impact of these vulnerabilities within the context of open-source ADS software projects:

\noindent\textbf{RQ1. What is the distribution of vulnerabilities across the source code of ADS projects?}

\noindent\textbf{Rationale:} This RQ aims to identify and characterize the types of vulnerabilities that are prevalent in ADS codebases. By understanding the distribution of these vulnerabilities, we can gain insights into potential coding practices or specific modules that may be more susceptible to defects.

\noindent\textbf{RQ2. How do vulnerabilities persist and evolve over time within the ADS project?}

\noindent\textbf{Rationale:} This RQ seeks to understand the life cycle of vulnerabilities from the perspective of developers and code reviewers. By tracking the process of vulnerabilities from being introduced to being resolved, we can assess the responsiveness and effectiveness of development teams in addressing security risks.

\noindent\textbf{RQ3. What is the tangible impact of identified vulnerabilities on the performance of ADS?}

\noindent\textbf{Rationale:} This RQ investigates the actual effects of vulnerabilities on the functionality of ADS. And by submitting selected vulnerabilities to the issue tracking system on GitHub and monitoring developer feedback, we aim to furthermore 
provide empirical evidence of the effectiveness of CodeQL in identifying security issues.

\subsection{Data Collection}
\begin{figure*}[h]
  \centering
  \includegraphics[width=1\textwidth]{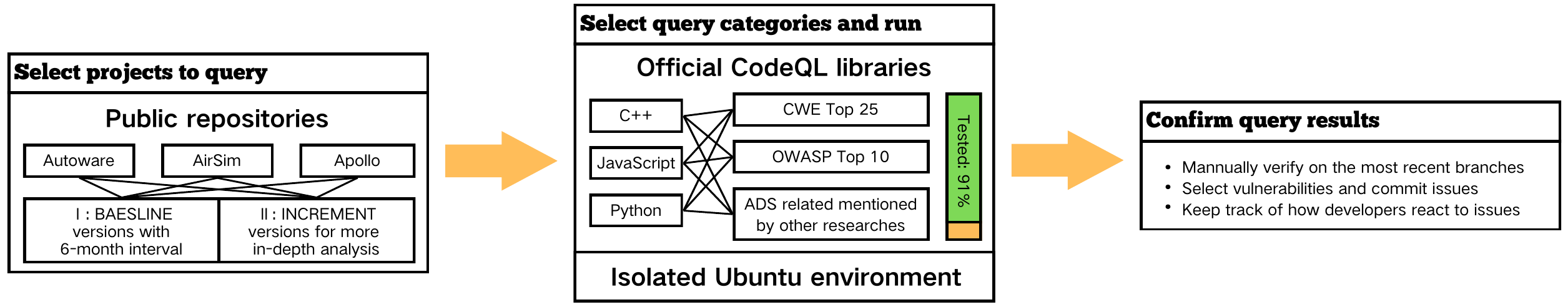}
  \caption{Overall Structure of the Data Collection Process}
  \label{research_design}
\end{figure*}

The overall process of data collection, as depicted in Figure \ref{research_design}, can be divided into three parts: selecting study cases, selecting query categories and subsequently using them to query, and confirming the results. They will be elaborated in the following parts.

\subsubsection{Select Projects and Versions}\label{Select_projects}

\begin{figure}[h]
  \centering
  \includegraphics[width=0.48\textwidth]{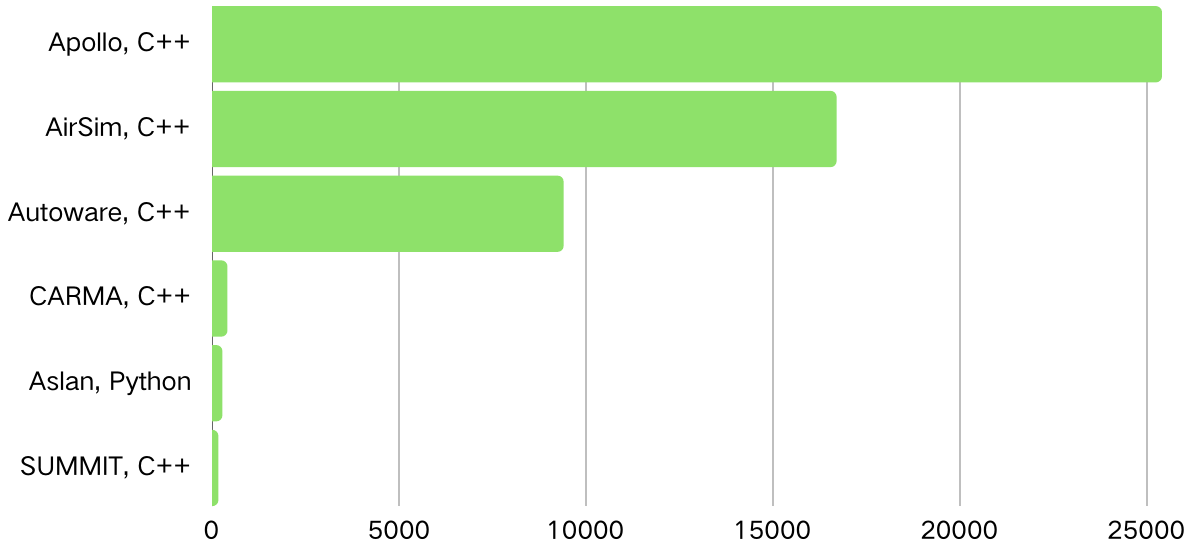}
  \caption{Well-known ADS Projects on GitHub, Their Main Language, and Number of Stars. Note. The data is obtained on 2025.1.21. And while there are many famous ADS-related repositories on GitHub, they do not contain the whole system.}
  \label{repo_stars}
\end{figure}

\textbf{Step 1: Select projects.} As shown in Figure \ref{repo_stars}, the number of \textbf{open-source} ADS projects is limited and their star counts have significant difference. Hence, our study focuses on three prominent open-source ADS projects with high GitHub star counts: Autoware, AirSim, and Apollo. These projects are recognized for achieving Level 4 autonomous driving capabilities~\citep{SAEJ3016_202104}, and have been selected for their representativeness and significant presence in the community. Firstly, Autoware is built upon the Robot Operating System (ROS), integrating a mature framework for robotics into the domain of autonomous driving. It has been selected by the U.S. Department of Transportation (USDOT) as a reference platform for smart transportation solutions. Secondly, AirSim offers a unique perspective by utilizing a game engine for realistic simulation of AVs, providing a virtual environment. It provides a realistic testing environment for developers, with access to a variety of sensor data and the ability to define behaviors through scripting. Lastly, Apollo stands out as a comprehensive and standalone autonomous driving software stack, encompassing an entire ADS ecosystem from perception to planning and control. It has also been integrated into educational curricula and real-world deployments. To guarantee the quality and representativeness of the code under analysis, we confine our study to the official releases of the projects from their public repositories. The approach ensures that we are examining code that has undergone the review and approval of project maintainers, thus reflecting a stable and tested state of the software.

\textbf{Step 2: Select the versions of projects.} We firstly examine the code of each project at intervals of approximately six months apart and our analysis in the initial phase involves a broad assessment across these selected intervals to identify patterns and trends in the code vulnerability profile. The interval is strategically chosen to reflect the dynamic evolution of open-source development and to capture the iterative refinements and enhancements made to the codebases over time. Based on the preliminary findings, we then select more versions for a more detailed analysis to refine our results and offer a more sophisticated perspective on the security of code. We track the version series beyond the six-month intervals and keep it continuous, delving into other historical versions to trace the evolution of vulnerabilities and the effectiveness of subsequent patches and fixes. Table \ref{table_project} shows the initially planned versions and the versions added to be tested at last. By adopting this methodical and adaptive sampling strategy, we aim to ensure that our data are both comprehensive and relevant, providing a solid foundation for our subsequent analysis and conclusions regarding the security of open-source ADS. 


\begin{table}[]
\caption{Release Versions to Be Tested of the Selected ADS}
\label{table_project}
\scalebox{1.0}{
\begin{tabular}{lll}
\hline
\begin{tabular}[c]{@{}l@{}}INITIALLY\\ PLANNED\end{tabular} &
  \begin{tabular}[c]{@{}l@{}}ADDITIONALLY\\ TESTED\end{tabular} &
  \begin{tabular}[c]{@{}l@{}}RELEASING\\ DATE\end{tabular} \\ \hline
Autoware-24.03      &               & 2024.03.12 \\
                    & Autoware-v1.0 & 2024.02.01  \\
Autoware-23.10      &               & 2023.10.24 \\
                    & Autoware-23.09 & 2023.09.05 \\ 
                    & Autoware-23.08 & 2023.08.29  \\
                    & Autoware-23.07 & 2023.07.11  \\
                    & Autoware-23.06 & 2023.06.12  \\
                    & Autoware-23.05 & 2023.05.29  \\ \hline
AirSim-1.8.0        &               & 2022.06.02 \\
AirSim-1.7.0        &               & 2022.01.12 \\
                    & AirSim-1.6.0  & 2021.08.24 \\
                    & AirSim-1.5.0  & 2021.03.15 \\
                    & AirSim-1.4.0  & 2021.01.09 \\ \hline
                    & Apollo-9.0.0  & 2023.12.18 \\
Apollo-9.0.0-alpha1 &               & 2023.08.02 \\
Apollo-8.0.0        &               & 2022.12.25 \\
                    & Apollo-7.0.0  & 2021.12.28 \\
                    & Apollo-6.0.0  & 2020.09.22 \\
                    & Apollo-5.0.0  & 2019.06.29 \\
                    & Apollo-3.0.0  & 2018.07.04 \\\hline
        \multicolumn{3}{p{210pt}}{Note. Autoware began to be restructured in late 2022 and Apollo-4.0.0 was not officially released and therefore not added to test.} \\
\end{tabular}
}
\end{table}

\subsubsection{Select Query Categories and Run} \label{difficulty}

\textbf{Step 1: Select representative languages.} Considering the overall code in selected ADS projects, the programming languages utilized are predominantly C++, followed by Python and JavaScript. Consequently, our analysis focuses on detecting vulnerabilities within these three languages, reflecting their prevalence in the project codebases. Given the distinct security measures inherent to each language, certain vulnerabilities are language-specific, and thus, some queries are tailored to target vulnerabilities unique to the language.

\begin{table*}[]
\caption{Overview of the Selected CodeQL Query Categories for ADS}
\scalebox{0.72}{
\begin{tabular}{llllllll}
\hline
\multirow{2}{*}{CATEGORY} &
  \multirow{2}{*}{NAME} &
  \multirow{2}{*}{\begin{tabular}[c]{@{}l@{}}CWE \\ TOP 25\end{tabular}} &
  \multirow{2}{*}{\begin{tabular}[c]{@{}l@{}}OWASP TOP \\ 10 RELATED\end{tabular}} &
  \multirow{2}{*}{TESTED} &
  \multicolumn{3}{l}{\begin{tabular}[c]{@{}l@{}}LANGUAGES \\ AVAILABILITY\end{tabular}} \\ \cline{6-8} 
        &                                                                                            &   &    &   & cpp & py & js \\ \hline
CWE-20 & Improper Input Validation                                                                  & Y & 01 & Y & Y   & Y  & Y  \\ \hline
CWE-78 & Improper Neutralization of Special Elements used in an OS Command (\textit{OS Command Injection}) & Y & 03 & Y & Y   & Y  & Y  \\ \hline
CWE-79 & Improper Neutralization of Input During Web Page Generation (\textit{Cross-site Scripting})       & Y & 03 & Y & Y   & Y  & Y  \\ \hline
CWE-89 & Improper Neutralization of Special Elements used in an SQL Command (\textit{SQL Injection})       & Y & 03 & Y & Y   & Y  & Y  \\ \hline
CWE-119 & Improper Restriction of Operations within the Bounds of a Memory Buffer                    & Y & N  & Y & Y   & N  & N  \\ \hline
CWE-120 & Buffer Copy without Checking Size of Input (\textit{Classic Buffer Overflow})                     & N & N  & Y & Y   & N  & N  \\ \hline
CWE-125 & Out-of-bounds Read                                                                         & Y & N  & Y & Y   & N  & N  \\ \hline
CWE-129 & Improper Validation of Array Index                                                         & N & N  & Y & Y   & N  & N  \\ \hline
CWE-134 & Use of Externally-Controlled Format String                                                 & N & N  & Y & Y   & N  & N  \\ \hline
CWE-190 & Integer Overflow or Wraparound                                                             & Y & N  & Y & Y   & N  & N  \\ \hline
CWE-191 & Integer Underflow (\textit{Wrap or Wraparound})                                                     & N & N  & Y & Y   & Y  & Y  \\ \hline
CWE-200 & Exposure of Sensitive Information to an Unauthorized Actor                                 & N & 02 & N & Y   & N  & N  \\ \hline
CWE-285 & Improper Authorization                                                                     & N & 01 & N & Y   & Y  & Y  \\ \hline
CWE-287 & Improper Authentication                                                                    & Y & 07 & N & N   & Y  & N  \\ \hline
CWE-290 & Authentication Bypass by Spoofing                                                          & N & 07 & Y & Y   & N  & N  \\ \hline
CWE-295 & Improper Certificate Validation                                                            & N & 06 & Y & Y   & Y  & Y  \\ \hline
CWE-311 & Missing Encryption of Sensitive Data                                                       & N & 02 & Y & Y   & N  & N  \\ \hline
CWE-313 & Cleartext Storage in a File or on Disk                                                     & N & 02 & Y & Y   & N  & Y  \\ \hline
CWE-319 & Cleartext Transmission of Sensitive Information                                            & N & 02 & Y & Y   & N  & N  \\ \hline
CWE-326 & Inadequate Encryption Strength                                                             & N & 02 & Y & Y   & Y  & Y  \\ \hline
CWE-327 & Use of a Broken or Risky Cryptographic Algorithm                                           & N & 02 & Y & Y   & Y  & Y  \\ \hline
CWE-352 & Cross-Site Request Forgery (CSRF)                                                          & Y & 03 & Y & N   & N  & Y  \\ \hline
CWE-367 & Time-of-check Time-of-use (TOCTOU) Race Condition                                          & N & N  & Y & Y   & N  & Y  \\ \hline
CWE-401 & Missing Release of Memory after Effective Lifetime                                         & N & N  & N & Y   & N  & N  \\ \hline
CWE-502 & Deserialization of Untrusted Data                                                          & Y & N  & Y & N   & Y  & Y  \\ \hline
CWE-611 & Improper Restriction of XML External Entity Reference                                      & N & 05 & Y & Y   & Y  & Y  \\ \hline
CWE-643 & Improper Neutralization of Data within XPath Expressions (\textit{XPath Injection})                & N & 03 & Y & N   & N  & Y  \\ \hline
CWE-676 & Use of Potentially Dangerous Function                                                      & N & N  & Y & Y   & N  & N  \\ \hline
CWE-704 & Incorrect Type Conversion or Cast                                                          & N & N  & Y & Y   & Y  & Y  \\ \hline
CWE-730 & Denial of Service                                                                          & N & N  & Y & N   & N  & Y  \\ \hline
CWE-732 & Incorrect Permission Assignment for Critical Resource                                      & N & 01 & Y & Y   & Y  & N  \\ \hline
CWE-776 & Improper Restriction of Recursive Entity References in DTDs (\textit{XML Entity Expansion})       & N & N  & Y & N   & Y  & Y  \\ \hline
CWE-787 & Out-of-bounds Write                                                                        & Y & N  & N & Y   & Y  & Y  \\ \hline
CWE-798 & Use of Hard-coded Credentials                                                              & Y & 07 & Y & N   & Y  & Y  \\ \hline
CWE-912 & Hidden Functionality                                                                       & N & N  & Y & N   & N  & Y  \\ \hline
CWE-915 & Improperly Controlled Modification of Dynamically-Determined Object Attributes             & N & N  & Y & N   & N  & Y  \\ \hline
CWE-918 & Server-Side Request Forgery (SSRF)                                                         & Y & N  & Y & N   & Y  & Y  \\ \hline
\multicolumn{8}{p{650pt}}{Note. The official CodeQL libraries has provided corresponding CWE number to the queries. However, due to the relatively broad categories in the OEWASP Top 10, the closest number association is additionally indicated.} \\
\end{tabular}%
}\label{table_query}
\end{table*}

\textbf{Step 2: Select query categories.} Table \ref{table_query} shows the specific query categories and their information. To ensure a comprehensive and targeted analysis, we have selected a set of queries based on the most critical vulnerabilities outlined in the \textit{2023 CWE Top 25 Most Dangerous Software Weaknesses}~\citep{2023_cwe_top25} and the \textit{OWASP Top 10: 2021}~\citep{owasp_top10_2021}. These lists represent a consensus within the cybersecurity community on the most pressing security risks faced by software applications, including those in the domain of ADS. Furthermore, we incorporate a review of the literature to identify vulnerabilities that are frequently cited in academic papers (especially the categories with \textbf{N} both under the \textbf{CWE TOP 25} and the \textbf{OWASP TOP 10 RELATED} columns in Table \ref{table_query}, but does not mean that other categories cannot be found in these papers) related to ADS~\citep{liang2016mlsa, liu2019edge, gupta2021deep, yang2023demystifying, mushtaq2017multilingual, kotenko2022static, xing2021toward, liu2020computing, feng2019understanding, li2022vulnerability, yaqoob2019autonomous}. This allows us to tailor our query categories to address the specific challenges and patterns observed within the ADS domain, ensuring that our analysis is not only broad but also relevant to the unique context of AVs. 

\textbf{Step 3: Ensure the validity of queries.} All queries are sourced directly from the official CodeQL libraries, which include a collection of \textit{supported queries}~\citep{codeql_supported_queries}. In the process of actually executing the experiment, these are queries outside the \textit{experimental/} directories in the CodeQL repository, which are noted as \textbf{Y} under the \textbf{TESTED} column in Table \ref{table_query}. The \textit{supported queries} have undergone testing and validation and are considered to provide accurate and useful results in most scenarios. They can be commonly used in production environments for code analysis, assisting developers in discovering and remediating security vulnerabilities and quality issues. In our study, we primarily employ these queries to ensure the reliability of our findings. This strategy aligns with best practices in software security research, leveraging the expertise and experience of the CodeQL development team and the wider cybersecurity community, while minimizing potential bias and ensuring a comprehensive detection of vulnerabilities across the diverse languages used in ADS development.

\begin{figure*}[h]
  \centering
  \includegraphics[width=1\textwidth]{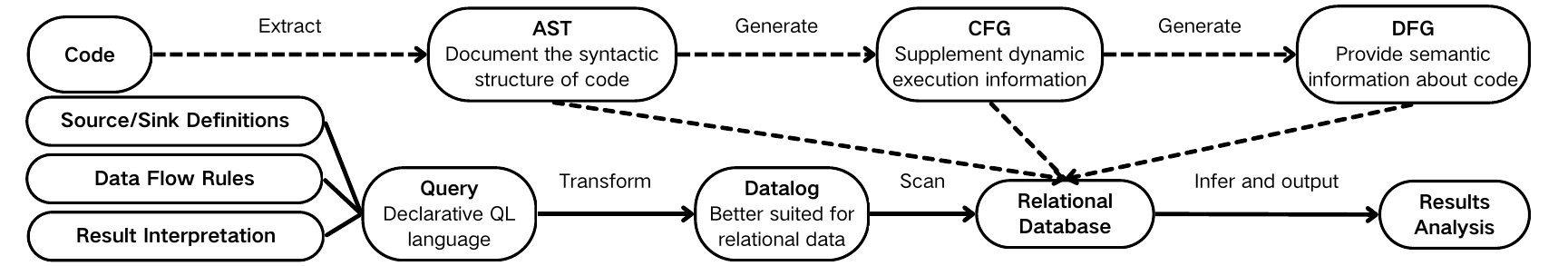}
  \caption{Workflow of CodeQL Vulnerability Detection}
  \label{codeql_process}
\end{figure*}

\textbf{Step 4: Run queries.} 
As depicted in Figure \ref{codeql_process}, CodeQL operates through a two-phase process: \textbf{database generation} and \textbf{query-based analysis}. In the first phase, the target codebase is parsed into a relational database that captures multi-layered semantic representations, including abstract syntax trees (ASTs), control-flow graphs (CFGs), and data-flow graphs (DFGs). During the process, interpreted languages will be analyzed directly through the source code, while compiled languages should be actually compiled to monitor compiler calls and use intermediate results. This database serves as a structured snapshot of the logic of code, enabling cross-file and cross-function analysis. The analysis phase employs CodeQL queries, written in the declarative QL language, to identify vulnerabilities or code patterns. A query usually includes source/sink definitions, data flow rules, and result interpretation, defining logical conditions over the database entities 
using classes and predicates. It will be transformed into Datalog language and then performed to scan the database to infer and output the final results. CodeQL’s strength lies in its ability to perform interprocedural analysis, recursively resolving function calls and variable aliases across the codebase. This enables precise detection of complex vulnerabilities
~\citep{avgustinov2016ql, de2007keynote, verbaere2007improve, github_codeql_scanning_docs, codeql_overview_docs}.

However, running queries for compiled languages is a complex and labor-intensive process that begins with building the project. This process demands meticulous planning and execution, as it involves creating isolated environments tailored to the specific requirements of each project version. Such environments must accurately replicate the original development settings to ensure the validity of analysis.

Firstly, we must identify and replicate the dependencies and prerequisites that are native to the original development context of the project. This often entails resolving broken or outdated links in the previous URL of the project, which may require extensive research to locate and replace missing resources.

Secondly, there is the challenge of dealing with libraries that are no longer supported or maintained. We must either find suitable replacements or, in some cases, reverse-engineer these libraries to ensure compatibility with the current development standards while maintaining the original functionalities.

Additionally, we must meticulously adhere to the project build rules and configurations. This involves not only installing the correct versions of compilers and tools but also ensuring that all custom build scripts and configurations are correctly applied. The ultimate goal is to pass all validation checks within the build rule set, which is crucial for verifying that the project is built correctly and the subsequent analysis can be conducted on a stable and reliable codebase.

\begin{figure}[h]
  \centering
  \includegraphics[width=0.48\textwidth]{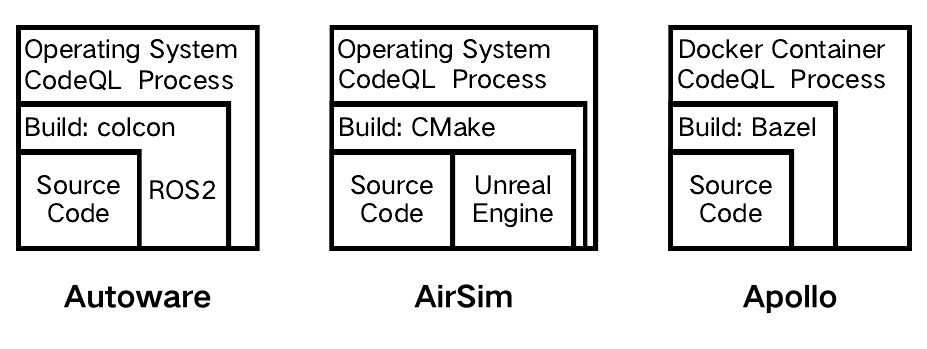}
  \caption{Relationships between Components in the Building Process}
  \label{build_process}
\end{figure}

During the general building process as shown in Figure \ref{build_process}, we utilize CodeQL to create databases for the code of the selected versions of each project. These databases are then subjected to analysis using our predefined suite of CodeQL queries, facilitating the detection and assessment of vulnerabilities. 


\subsubsection{Confirm Query results} \label{Comfirm}

\textbf{Step 1: Manually validate.} As the version iteration goes on, we validate the vulnerabilities identified between the tested versions. This aims to test the validity and accuracy of the results, confirm whether the vulnerabilities have been resolved by the development community, and check if the features in the data are caused by solving problems while introducing others. To quantify the inter-rater reliability of our manual validation process, we employed Cohen's Kappa coefficient~\citep{cohen1960coefficient}. The first and second authors independently validated a subset of identified vulnerabilities, and the Kappa coefficient was calculated to evaluate the consistency of their judgments. We focused on directories that were reported to have a denser distribution of vulnerabilities and conducted verification and comparison for each category by comparing the same file in different versions and following CodeQL analysis steps to check from \textit{source} to \textit{sink} as it defined. The resulting Kappa value was 0.8213, indicating a substantial agreement between the researchers. This high level of agreement suggests that our manual validation process is reliable and consistent.

\textbf{Step 2: Submit issues.} 
After manual validation, we selected a set of representative vulnerabilities that are characterized by their frequency of reporting. Due to its inherent mechanism, CodeQL identifies and reports vulnerabilities all together, which include both general bugs and security issues related to CVE (Common Vulnerabilities and Exposures). As a result of code clone, similar vulnerabilities can distribute widely and be reported many times separately~\citep{mo2023comprehensive}. However, higher severity levels are also indicated because of their frequency of occurrence. Hence, after checking their existence in the latest developing version, similar vulnerabilities are wrapped as a whole and then systematically reported as issues to the projects that interact more with developers, namely Autoware and Apollo. The development team of AirSim has not responded to the developer in over six months.

\textbf{Step 3: Keep track on reactions.} We track the responses and feedback of the development teams once the issues are submitted. This involves monitoring the communication as well as the speed at which the reported vulnerabilities are resolved. By observing their engagement and response, we are able to assess whether the reported vulnerabilities are indeed severe and have been overlooked in the development process. 

\subsection{Data Analysis}\label{Data_Analysis}

The outputs produced by CodeQL are in the .sarif (Static Analysis Results Interchange Format) format, encapsulating the results in a JSON structure. To answer RQ1 and RQ2 by analyzing the output, we develop and employ custom script programs (also undergone same test procedures in Section \ref{Comfirm}) that parse the .sarif files, focusing on two critical dimensions: the type and location of the identified vulnerabilities. These scripts are tailored to extract and categorize the relevant information from the JSON structure, enabling us to map and quantify the distribution of vulnerabilities within the codebases of the selected ADS projects.

As for RQ3, the quality and timeliness of the developers feedback are considered as indicators of the gravity with which the project considers the reported issues. If the development teams demonstrate a prompt and thorough response, it may suggest that the vulnerabilities are recognized as significant and the project has effective mechanisms for addressing security concerns. Conversely, delayed or missing responses could indicate that the issues are not perceived as urgent or considered solvable within the project. 

Furthermore, the process also provides empirical evidence of the value of using CodeQL for vulnerability detection. By comparing the issues identified through CodeQL with the feedback and subsequent resolution of these issues, we can gauge the effectiveness of CodeQL in uncovering vulnerabilities that could potentially impact the performance of ADS. This comparison sheds light on the practical utility of static code analysis tools like CodeQL in the context of open-source ADS development, reinforcing the importance of integrating such tools into the development life cycle to ensure the security and reliability of these complex systems. 

\section{Results And Analysis} \label{Results}
We present the results and corresponding analysis of the three RQs in the following. The raw data of the query results are available online~\citep{CodeqlRes}.
\subsection{Distribution of Vulnerabilities (RQ1)}
\subsubsection{Distribution of Vulnerabilities in Different Categories}
\begin{figure*}[h]
  \centering
  \includegraphics[width=0.8\textwidth]{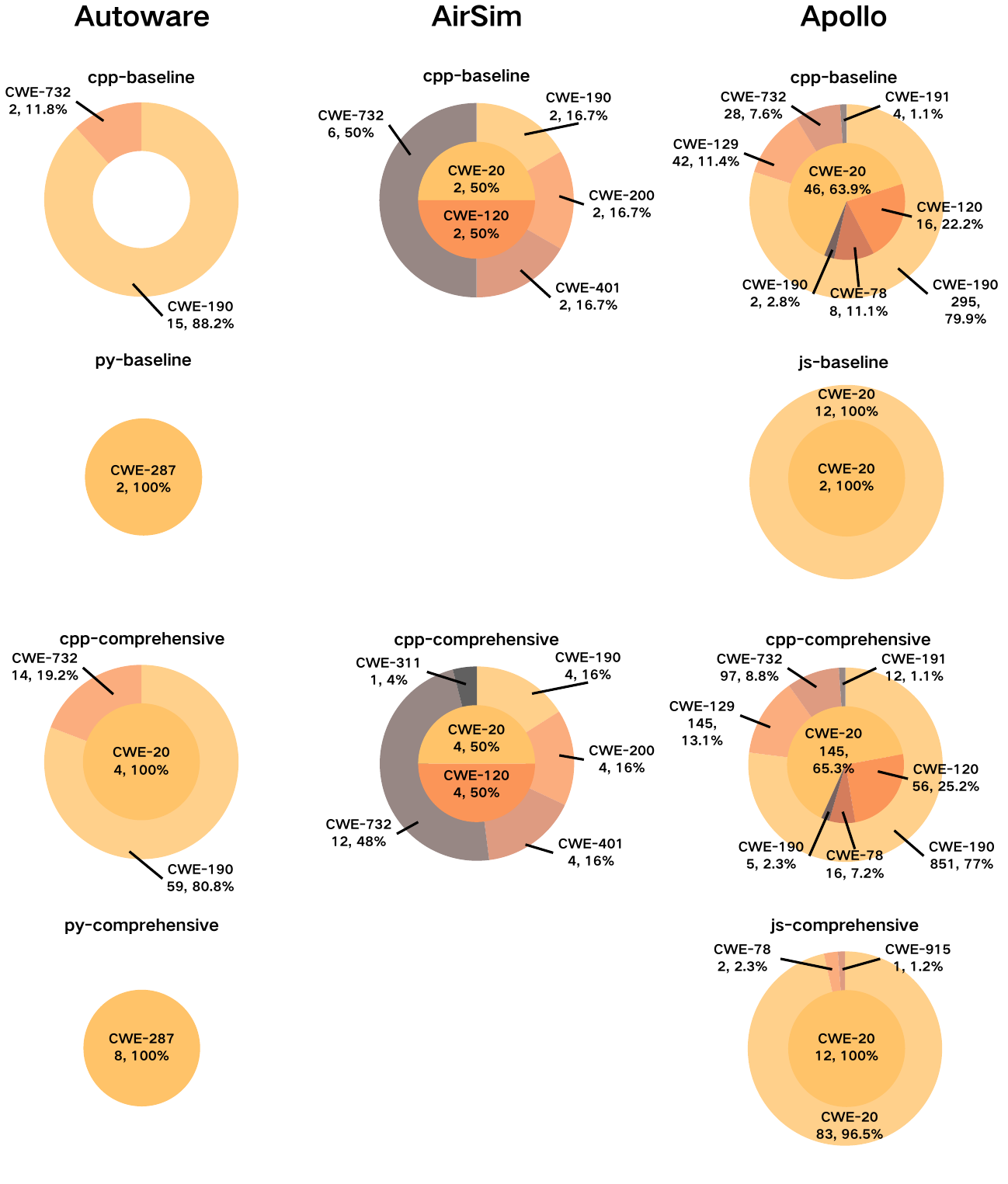}
  \caption{Distribution of CWE Vulnerabilities (RQ1). Note. The inner circle represents the error level, and the outer circle represents the warning level.}
  \label{percentage_distribution}
\end{figure*}

 We add and exhibit same type of vulnerabilities from all tested versions together. As shown in Figure \ref{percentage_distribution}, 37 queries were set to test every project, and 12 types of vulnerabilities were detected. Across the projects, the most commonly identified errors and warnings were CWE-190 (Integer Overflow or Wraparound) and CWE-20 (Improper Input Validation). Meanwhile, vulnerabilities such as CWE-191 (Integer Underflow, Wrap or Wraparound) and CWE-311 (Missing Encryption of Sensitive Data) were largely absent from the results. And despite the comprehensive set of categories predefined for detection, substantial types of vulnerabilities were not detected. In this section, the focus has been on vulnerabilities intrinsic to ADS projects themselves, excluding those introduced through third-party libraries. This exclusion is intentional, as third-party vulnerabilities could skew the data and distract the primary objective of assessing the security of the ADS codebase. 

In Autoware, warnings were predominantly found in C++ code, with CWE-190 and CWE-732 accounting for 88.2\% and 11.8\% of the errors, respectively, in the baseline testing phase. In the increment phase, these percentages remained similar. Python code only contained CWE-287 errors, with 100\% occurrence in both phases. 


In the AirSim project, the error distribution across the baseline and increment testing phases showed a consistent distribution. In C++ code, CWE-20 and CWE-120 were the primary error categories, each constituting 50\% of the errors in both phases. For warnings, CWE-732, CWE-190, CWE-200, and CWE-401 were identified, with CWE-732 accounting for 50\% in the baseline phase and 48\% in total. CWE-190, CWE-200, and CWE-401 each made up 16.7\% of the warnings in the baseline phase, while CWE-190 and CWE-200 maintained the same percentage at last, with CWE-401 also constituting 16\%. Notably, CWE-311 was introduced as a new category of warning in the increment phase, representing 4\% of the warnings. In Python code, no errors or warnings were detected in either phase. 

In the Apollo project, the baseline testing phase revealed that in C++ code, CWE-20 and CWE-120 were the most prevalent error categories, comprising 63.9\% and 22.2\% of the errors, respectively. CWE-78 and CWE-190 accounted for 11.1\% and 2.8\% of the errors, respectively. In the warning category, CWE-190 was the most common, representing 79.9\% of the warnings, followed by CWE-129, CWE-732, and CWE-191. Observed with the increment phase, which included additional versions, the distribution of errors in C++ code remained similar, with CWE-20 and CWE-120 still dominating at 65.3\% and 25.2\%, respectively. The percentages of CWE-78 and CWE-190 errors decreased slightly to 7.2\% and 2.3\%, respectively. The warning category continued to be dominated by CWE-190, with a slight decrease to 77\%, and the other categories remained relatively stable. In JavaScript code, CWE-20 was the only error category in the baseline phase, representing 100\% of the errors and warnings. Apollo-9.0.0 introduced a few other vulnerability types in the increment phase but CWE-20 remained as the dominant type. An error was detected in the Python code of Apollo-9.0.0 but it only participated in document generation, not in real driving process. So it was excluded from statistics as well. The following parts will keep this choice. 

\textbf{Interpretation: }The frequently detected categories indicate a consistent pattern of vulnerabilities that requires focused remediation efforts. While different projects are established based on different viewpoints with different building approaches, CWE-190 (Integer Overflow or Wraparound) and CWE-20 (Improper Input Validation) were commonly detected, along with other vulnerabilities. They potentially compromise the security and reliability of the system, as CWE-190 can lead to unpredictable system behavior or even crashes and CWE-20 can allow unauthorized or malicious data to be processed by the system. In the domain of ADS, this could manifest as erratic vehicle control responses or malfunctions in the sensor data processing algorithms. All of these vulnerabilities underscore the critical need for robust input handling and data processing mechanisms within ADS. Addressing these issues requires a combination of rigorous coding practices, comprehensive testing, and the implementation of safeguards to detect and mitigate the impact of such vulnerabilities.

\subsubsection{Distribution of Vulnerabilities in Different Modules}
In this part, we indicate the specific number of the three vulnerabilities in the form of (error, warning, note) triples with them averaged based on the number of tested versions, giving information about what a ``typical'' project looks like. As previously discussed in Section \ref{Select_projects}, these projects have some unique factors. However, it turns out that the dense distribution observed in directories such as \textbf{perception/} for Autoware, \textbf{MavLinkCom/} for AirSim, and \textbf{lidar/} for Apollo suggests several common features and the detailed functions of each module are listed in Table \ref{table_Modules}. The vulnerability statistics also exclude those introduced through third-party libraries in this part, focusing on the intrinsic security of the project codebases. 



\begin{table*}[]
\caption{The Severity of the Vulnerability under Different Modules in the Autoware Project (RQ1)}
\scalebox{1.0}{
\begin{tabular}{llll}
\hline
\begin{tabular}[c]{@{}l@{}} RANGE\end{tabular} & \multicolumn{2}{l}{DIRECTORY} & \begin{tabular}[c]{@{}l@{}}SEVERITY\end{tabular} \\ \hline
BASELINE  & universe/ &                      & (1, 4, 0) \\
         &           & perception/          & (0, 3, 0) \\
         &           & planning/            & (1, 0, 0) \\
         &           & system/              & (0, 1, 0)  \\
         & external/ &                      & (0, 5, 0) \\
         &           & rtklib\_ros\_bridge/ & (0, 4, 0) \\
         &           & eagleye/             & (0, 1, 0) \\ \hline
COMPREHENSIVE & universe/ &                      & (1, 3.5, 0) \\
         &           & perception/          & (0, 2.5, 0) \\
         &           & planning/            & (1, 0, 0) \\
         &           & system/              & (0, 1, 0)  \\
         & external/ &                      & (0, 4.5, 0) \\
         &           & rtklib\_ros\_bridge/ & (0, 3.5, 0) \\
         &           & eagleye/             & (0, 1, 0)  \\
         & sensor\_component/ & external/velodyne\_vls    & (0.8, 0, 0)  \\\hline
         \multicolumn{4}{p{390pt}}{Note. ``COMPREHENSIVE'' means the results combine the data of the initial and incremental phases.} \\
\end{tabular}%
   }
\label{table_Autoware}
\end{table*}

In the Autoware project (Table \ref{table_Autoware}), both phases revealed a significant distribution of vulnerabilities within specific second-level directories. The \textbf{universe/} directory presented the highest severity across them. The \textbf{external/} directory also showed a stable distribution of vulnerabilities, with warnings of 5 and 4.5 in instances and the \textbf{rtklib\_ros\_bridge/} directory maintained the largest share of them. The sub-directories such as \textbf{perception/}, \textbf{planning/}, and \textbf{system/} had less and relatively even vulnerabilities, and the \textbf{eagleye/} directory remained at the lowest end with a warning. And it is worth noting that \textbf{sensor\_component/} had some vulnerabilities originally but they were fixed in later development.


\begin{table*}[]
\caption{The Severity of the Vulnerability under Different Modules in the AirSim Project (RQ1)}
\scalebox{1.0}{
\begin{tabular}{llll}
\hline
RANGE & \multicolumn{2}{l}{DIRECTORY} & \begin{tabular}[c]{@{}l@{}}SEVERITY\end{tabular} \\ \hline
BASELINE  & MavLinkCom/ &              & (1, 5, 0) \\
         &             & src/         & (1, 4, 0) \\
         &             & MavLinkTest/ & (0, 1, 0)  \\
         & DroneShell/ & include /    & (1, 0, 0) \\
         & AirLib/     & eigen3/      & (0, 1, 0)  \\ \hline
COMPREHENSIVE & MavLinkCom/ &              & (1, 5.2, 0) \\
         &             & src/         & (1, 4.2, 0) \\
         &             & MavLinkTest/ & (0, 1, 0)  \\
         & DroneShell/ & include /    & (1, 0, 0) \\
         & AirLib/     & eigen3/      & (0, 1, 0)  \\ \hline
\end{tabular}
}
\label{table_AirSim}
\end{table*}

In the AirSim project (Table \ref{table_AirSim}), the vulnerability analysis highlighted certain first-level directories with a higher concentration of vulnerabilities. Notably, the \textbf{MavLinkCom/} directory had the most errors and warnings of (1, 5) initially and (1, 5.2) eventually. The \textbf{src/} subdirectory under \textbf{MavLinkCom/} also showed considerable density of (1, 4) initially and (1, 4.2) eventually and the \textbf{include/} subdirectory had an error in both testing phases. The \textbf{MavLinkTest/} and  \textbf{eigen3/} subdirectories consistently had a warning, showing a minimal presence of severe vulnerabilities.  


\begin{table*}[]
\caption{The Severity of the Vulnerability under Different Modules in the Apollo Project (RQ1)}
\scalebox{1.0}{
\begin{tabular}{lllll}
\hline
RANGE & \multicolumn{3}{l}{DIRECTORY} & \begin{tabular}[c]{@{}l@{}}SEVERITY\end{tabular} \\ \hline
BASELINE  & modules/      &               &             & (15, 113.5, 0) \\
         &               & drivers/      &             & (3, 58, 0) \\
         &               &               & lidar/      & (3, 54, 0) \\
         &               &               & smartereye/ & (0, 2, 0)  \\
         &               &               & camera/     & (0, 1, 0)   \\
         &               &               & microphone/ & (0, 1, 0)   \\
         &               & localization/ &             & (3, 18, 0) \\
         &               & perception/   &             & (1, 18, 0) \\
         &               & dreamview/    &             & (5, 6, 0)  \\
         &               & bridge/       &             & (3, 3, 0)  \\
         &               & others/       &             & (0, 10.5, 0) \\
         & third\_party/ & rtklib/       &             & (22, 84, 0) \\ \hline
COMPREHENSIVE & modules/      &               &             & (16.9, 113.9, 0) \\
         &               & drivers/      &             & (7.9, 63.1, 0) \\
         &               &               & lidar/      & (2, 29.1, 0) \\
          &               &               & gnss/ & (5.4, 18.3, 0)   \\
          &               &               & velodyne/ & (0, 12.9, 0)   \\
         &               &               & smartereye/ & (0, 1.1, 0)  \\
         &               &               & camera/     & (0.1, 0.7, 0)   \\
         &               &               & microphone/ & (0, 1, 0)  \\ 
          &               &               & hesai/ & (0.4, 0, 0)   \\
         &               & localization/ &             & (1.9, 17.3, 0) \\
         &               & perception/   &             & (1.1, 16.3, 0) \\
         &               & dreamview/    &             & (2.7, 5.7, 0)  \\
         &               & dreamview\_plus/    &             & (0.6, 5.7, 0)  \\
         &               & bridge/       &             & (2.7, 2.1, 0) \\
         &               & others/       &             & (0.6, 9.4, 0)  \\
         & third\_party/ & rtklib/       &             & (14.9, 54.3, 0) \\
         & cyber/        & io/           &             & (1.3, 0, 0)   \\ \hline
         \multicolumn{5}{p{390pt}}{Note. We divided the drivers/ directory into smaller parts for it containing considerable vulnerabilities.} \\
\end{tabular}
}
\label{table_Apollo}
\end{table*}

In the Apollo project (Table \ref{table_Apollo}), \textbf{velodyne/} and \textbf{hesai/} were later merged into \textbf{lidar/}, and \textbf{gnss/} was eventually separated from \textbf{drivers/}, operating independently in \textbf{third\_party/}. Meanwhile, \textbf{dreamview\_plus/} was newly added in Apollo-9.0.0. We kept them in original place to ensure data integrity. Revealing notable severity in both phases, \textbf{modules/} presented the most vulnerabilities, with (15, 113.5) initially and (16.9, 113.9) eventually and the included \textbf{drivers/} showed the highest severity in both phases. And as it reached a much higher density than any other directories, we further explored the module and found that the code in \textbf{lidar/} contained the most vulnerabilities. Additionally, \textbf{third\_party/rtklib/} had substantial vulnerabilities of (22, 84) and (14.9, 54.3) in two phases. The \textbf{cyber/io/} directory, introduced in Apollo-5.0.0 and Apollo-9.0.0, had less errors of 1.3. 

\textbf{Interpretation: }In brief, the Autoware project shows a consistent vulnerability pattern with higher severity in modules like \textbf{Perception} and \textbf{Planning}. In AirSim, the \textbf{MavLinkCom} stands out with the highest severity level. Analysis of Apollo points to the \textbf{modules/} and \textbf{third\_party/} directories, particularly \textbf{Lidar} and \textbf{rtklib}, as areas with significant vulnerability concentrations. By associating the results mentioned above with the functionality of the modules (Table \ref{table_Modules}), it can be found that vulnerabilities appear more frequently in the perception-related modules, including processing sensor data, especially LiDAR data, Real-time kinematic positioning, and telemetry and control data transfer. Meanwhile, directories with higher severity tend to contain more complex code such as multiple programming languages and involve extensive hardware interactions, all of which can increase the likelihood of vulnerabilities. The diversity of code and interactions increases the attack surface, making these areas more prone to security issues that require a security focus. These perception-related modules suggest that the code may have consistent vulnerabilities that need to be addressed.

In addition, although 37 queries were set to test every project, only 12 types of vulnerabilities were eventually detected. Even though a high number of undetected categories and a decrease in the absolute number of detected vulnerabilities exist, the identified vulnerability distribution from our scan of the source code remains similar to the conclusions drawn from commits~\citep{garcia2020comprehensive}. This finding suggests that the ADS projects have implemented effective countermeasures against these specific vulnerabilities, and there is still room for further improvement.


\begin{table*}[]
\caption{Functionality of Reported Modules}
\scalebox{1.0}{
\begin{tabular}{p{0.18\columnwidth}p{0.35\columnwidth}p{1.3\columnwidth}}
\hline
PROJECT & MODULES          & FUNCTIONALITY                                       \\ \hline
Autoware & Perception        & Processes sensor data for object detection, classification, and tracking   \\ \cline{2-3} 
        & Planning         & Path planning and decision-making  \\ \cline{2-3} 
        & System           & System integration and coordination                \\ \cline{2-3} 
        & External         & Interface with external systems or services        \\ \cline{2-3} 
         & rtklib ROS Bridge & Real-time kinematic positioning (RTK) related to improve positioning accuracy \\ \cline{2-3} 
        & Eagleye          & Visual processing                                  \\ \cline{2-3} 
        & Sensor Component & Handling and preprocessing of sensor data          \\ \cline{2-3} 
        & Velodyne VLS     & Processing data of high-resolution LiDAR sensors   \\ \hline
AirSim   & MavLinkCom        & MavLink communication protocol for telemetry and control data transfer     \\ \cline{2-3} 
         & DroneShell        & Command-line interaction or script execution for drones                    \\ \cline{2-3} 
         & AirLib            & Core library for simulating the physical and visual behavior of drones     \\ \cline{2-3} 
        & Eigen3           & Support for mathematical calculations              \\ \hline
Apollo  & Modules          & Encompasses various autonomous driving modules     \\ \cline{2-3} 
        & Drivers          & Hardware drivers, such as sensor drivers           \\ \cline{2-3} 
        & Lidar            & Processing for LiDAR sensor data                   \\ \cline{2-3} 
        & Smartereye       & Intelligent vision system                          \\ \cline{2-3} 
        & Camera           & Processing for camera sensor data                  \\ \cline{2-3} 
        & Localization     & Vehicle positioning                                \\ \cline{2-3} 
        & Perception       & Environmental perception                           \\ \cline{2-3} 
        & Dreamview        & User interface or visualization tool               \\ \cline{2-3} 
        & Bridge           & Data bridging between different systems            \\ \cline{2-3} 
        & Others           & Miscellaneous functionalities                      \\ \cline{2-3} 
        & Third-Party      & Third-party libraries, such as RTK for positioning \\ \cline{2-3} 
        & Cyber IO         & Data input/output and communication protocols      \\ \hline
\end{tabular}
}
\label{table_Modules}
\end{table*}

\FrameSep=3pt\FrameRule=0.5pt\begin{framed}
\noindent \textbf{Answer to RQ1:} The distribution of vulnerabilities in ADS projects reveals a concentration of \textbf{CWE-190} and \textbf{CWE-20} errors and warnings. And 
complex modules with \textbf{perception-related} functionality tend to contain more vulnerabilities.
\end{framed}

\subsection{Life Cycle of Vulnerabilities (RQ2)}
\begin{figure*}[h]
  \centering
  \includegraphics[width=0.82\textwidth]{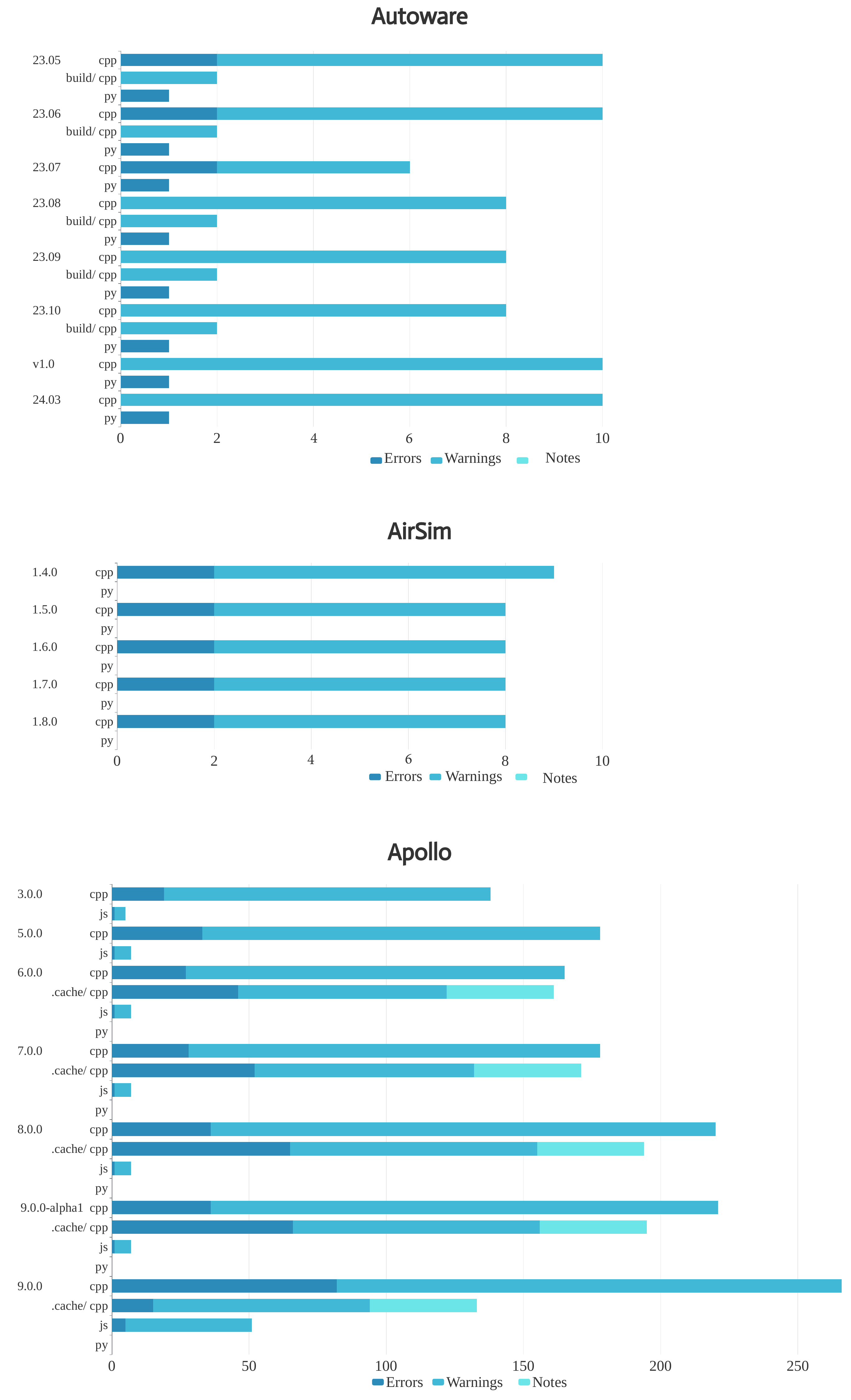}
  \caption{Distribution of Vulnerabilities in Different Languages (RQ2). Note. The version of python language in Apollo-3.0.0 and 5.0.0 is stale and therefore not tested.}
  \label{volume_distribution}
\end{figure*}

After confirming in different versions, the results of the sampled code across various projects, as are shown in Figure \ref{volume_distribution}, have been observed that once vulnerabilities are introduced, they tend to persist through version iterations. Moreover, the introduction of vulnerabilities through the use of third-party libraries during the build process is a common occurrence, and it makes a difference when newer versions of third-party libraries are used.

Across the two testing stages for Autoware, the number of vulnerabilities showed a relative constancy. From Autoware-23.08 to Autoware-23.10, there were 8 warnings in C++ and 1 error in Python consistently in each version. In the more recent Autoware-v1.0, the distribution changed a little, similarly with 10 warnings in C++ and 1 error in Python. And the vulnerabilities in \textbf{build/} disappeared, meaning that they were fixed by third-party libraries. Similar operations were completed by Autoware-23.07, solving vulnerabilities both from developing code and third-party libraries.


In the AirSim project, the vulnerabilities across the assessed versions indicate a similar and consistent distribution pattern. Starting with AirSim-1.4.0, which was added to the analysis, there were 2 errors and 7 warnings in C++ code, with no reports about Python. This pattern persisted through AirSim-1.5.0 and AirSim-1.6.0, with similar counts of 2 errors and 6 warnings in C++, and none in Python. The initial testing of AirSim-1.7.0 and AirSim-1.8.0 showed no change in the number of vulnerabilities, with 2 errors and 6 warnings in C++ continuing to be the case, and Python remaining clear of detected vulnerabilities. 

Noticeable changes appeared basically every two versions in Apollo and we especially focused on versions began at Apollo-8.0.0 as they turned out to present a more obvious and dynamic change. In Apollo-8.0.0 and Apollo-9.0.0-alpha1, the distribution of vulnerabilities across different severity levels remained relatively consistent. However, in the more recent Apollo-9.0.0 version, there is a significant increase in the number of error-level vulnerabilities within the codebase. Additionally, it is observed that the number of vulnerabilities introduced in the \textbf{.cache/} directory, which often results from the use of third-party libraries during the building process, has decreased notably in the Apollo-9.0.0 version, with errors dropping by 51 and warnings dropping by 11.

\textbf{Interpretation: }Most vulnerabilities often go unnoticed or unaddressed and the persistent security issues have not been fully resolved in most cases. In Autoware, the pattern of vulnerabilities distribution remained consistent for more than half a year, in AirSim the number is 15 months, and in Apollo, while the distribution changed after a year, vulnerabilities increased. The development team has introduced new code and made changes that resulted in a higher number of issues that could potentially lead to security breaches or system failures if not addressed efficiently.

And as to deal with third-party libraries, vulnerabilities both in Autoware and Apollo show a reduction through version iterations indicating that the project has likely taken steps to improve the security of their build processes. A typical remediation strategy involves updating to newer versions of these libraries, which may include security patches and fixes for known issues. Enhancing the security configurations and implementing more rigorous security checks during the build phase also helps.


\FrameSep=3pt\FrameRule=0.5pt\begin{framed}
\noindent \textbf{Answer to RQ2:} The vulnerabilities tend to persist across version iterations, usually for \textbf{more than 6 months}. Notably, vulnerabilities can also be introduced during the build process and can be effectively addressed by updating third-party libraries.
\end{framed}

\subsection{Results of Manual Validation and Feedback from Developers (RQ3)}  \label{examples}
In the subsequent phase of vulnerability detection, the confirmed vulnerabilities that appeared frequently and were most confident to be true positives were submitted as issues, with 2 about Autoware and 4 about Apollo, to the respective projects. As the outcomes, 2 identified vulnerabilities about Autoware, CWE-190 (Integer Overflow or Wraparound), were acknowledged and resolved by the development teams in the first month~\citep{WebPage2024}. Listing \ref{code-2} shows the details of code. The resolution of the identified vulnerabilities, which were classified as a warning in the CodeQL reports, attests to the commendable detection capabilities of the tool. The warning, though not critical error, signify potential security risks that could escalate under certain conditions.

\textbf{Interpretation: } The rapid response time indicates an acknowledgment of the potential impact these vulnerabilities could have on the integrity and functionality of system. Given the dynamic nature of software development, with ongoing code updates and the associated evolution of potential vulnerabilities, the feasibility of a one-time exhaustive validation of all detected vulnerabilities is limited. The resource-intensive nature of manual verification further compounds this challenge. However, this also suggests that there may be vulnerabilities that have yet to be addressed. Therefore, it is imperative to incorporate regular CodeQL scans into the development process to ensure ongoing vigilance against emerging security threats.

The reactions from developers serve as a proof of the effectiveness of CodeQL and an objective judgment of security to ADS. We provide examples of its identification reports below. 

\begin{lstlisting}[language=C++, caption=Apollo CWE-78, label=code-1]
...
  const std::string home = cyber::common::GetEnv("HOME"); //call to GetEnv
  *scenario_resource_path = home + FLAGS_resource_scenario_path; //call to operator+ then operator= output argument
...
  GetScenarioResourcePath(&directory_path); //GetScenarioResourcePath output argument
  directory_path = directory_path + scenario_set_id;
  if (!cyber::common::PathExists
            (directory_path)) {
    AERROR << "Failed to find scenario_set!";
    return;
  }
  std::string command = "rm -fr " + directory_path; //*directory_path then call to operator+

  if (std::system(command.data()) != 0) { //*call to data
    AERROR << "Failed to delete scenario set directory for: "
           << std::strerror(errno);
    return;
...
\end{lstlisting}
In Listing \ref{code-1}, using user-supplied data in an OS command, without neutralizing special elements, can make code vulnerable to command injection.

\begin{lstlisting}[language=C++, caption=Autoware CWE-190, label=code-2]
...
  if (roi.x_offset + roi.width > width) {
    roi.width = width - roi.x_offset;
  }
...
\end{lstlisting}
In Listing \ref{code-2}, writing `if (a+b>c) a=c-b' incorrectly implements `a = min(a,c-b)' if `a+b' overflows. This integer overflow is the root cause of the buffer overflow in the SHA-3 reference implementation (CVE-2022-37454).

\begin{lstlisting}[language=C++, caption=Apollo CWE-191, label=code-3]
...
    size_t lap = sp_laps_checker_->GetLap();
...
    if (data_type == DataType::MAP_CHECKOUT) {
      if (is_reached) {
        loop_result->set_loop_num(
            static_cast<double> (sp_conf_-> laps_number_additional));
      } else {
        loop_result->set_loop_num(
            lap - sp_conf_->laps_number >= 0 //Unsigned subtraction can never be negative.
                ? static_cast<double>(lap - sp_conf_->laps_number)
                : 0.0);
      }
    }
...
\end{lstlisting}
In Listing \ref{code-3}, it does not make sense to compare whether the result is greater than or equal to 0, because the result of this expression will be non-negative anyway. A subtraction with an unsigned result can never be negative. Using such an expression in a relational comparison with `0' is likely to be wrong.

The results also suggested that the adoption of CodeQL in the development of ADS remains limited. One reason for this is that CodeQL is a relatively new tool, and as emphasized in Section \ref{difficulty}, its use requires a certain level of expertise and effort. Development teams may face resource and time constraints that limit their ability to implement and maintain CodeQL in their workflows. Furthermore, echoed by~\citep{fischer2023effectiveness}, even though several studies have confirmed the effectiveness of CodeQL, developers still harbor doubts about the accuracy and reliability of static analysis tools in general. This skepticism may stem from concerns about false positives and the potential for these tools to generate a large number of alerts that require manual verification. Moreover, the lack of widespread integration of this tool into development practices~\citep{fischer2023effectiveness} means that there is no established industry standard or best practice for its use. This absence of a common approach further discourages developers from adopting CodeQL, as they may feel uncertain about how to best integrate it into their existing workflows and how to interpret and act on its findings. As a result, the potential benefits of using CodeQL to enhance the security of ADS and other projects are not fully realized, highlighting the need for more education, support, and standardization in this area.

When it comes to ADS, it is necessary to acknowledge that even a small number of detected vulnerabilities, though minor and few in the vast expanse of the codebase, can lead to significant harm due to the system's direct impact on human safety. Therefore, the security standard in ADS must be exceptionally high to ensure public safety and trust. Our study shows that while the majority of the code in ADS projects may be secure, and the system certainly works in most cases, the potential for severe consequences from an unnoticed vulnerability still exists and needs more attention. This includes more efficient detection approaches and more standardized develop and problem-resolving measures.


\FrameSep=3pt\FrameRule=0.5pt\begin{framed}
\noindent \textbf{Answer to RQ3:} Upon submission of findings as issues on GitHub, we observed an active response from the development community. This not only demonstrates the \textbf{tangible impact} of the identified vulnerabilities but also reflects the \textbf{effectiveness} of CodeQL in uncovering actionable security issues. 
\end{framed}



\section{Implications} \label{Discussion}

Considering the number of stars obtained by repositories and the universality of their application, though limited ADS projects were selected, our conclusions should be able to be partially generalized to other ADS projects, particularly those with similar characteristics (e.g., multi-language codebases, open-source development). The implications of this research are multifaceted, offering valuable insights for developers, researchers, and policymakers in the realm of ADS.

\subsection{Implications for Developers}
The findings of this study underscore the utility of CodeQL in the development of ADS. By integrating static code analysis tools like CodeQL for security weakness detection into the development process, developers can automate the detection of potential vulnerabilities, thereby enhancing the quality of the code. It is recommended for developers to: 
\textbf{Adopt CodeQL in Development Workflows:} Integrate CodeQL into continuous integration (CI) pipelines to automatically scan codebases for vulnerabilities during each build process.
\textbf{Focus on High-Risk Modules:} Separating those most vulnerable modules from others is advisable for both testing and management. Given the prevalence of vulnerabilities in perception-related modules (e.g., LiDAR data processing), developers should prioritize these areas for more rigorous testing and code reviews.
\textbf{Regularly Update Third-Party Libraries:} Ensure that third-party libraries are kept up-to-date to mitigate the introduction of new vulnerabilities. This practice can significantly reduce the number of vulnerabilities introduced during the build process, as newer versions often include patches and fixes for known security issues.



\subsection{Implications for Researchers}

The study highlights the ongoing challenges associated with static code analysis, including the potential for false positives and related concerns. Researchers can contribute to the field by:
\textbf{Improving Query Accuracy:} Refine and expand the query categories used in CodeQL to improve both the accuracy and recall of vulnerability detection. This includes developing more sophisticated algorithms to minimize the reliance on manual validation.
\textbf{Investigating Vulnerability Persistence:} Study the reasons behind the persistence of vulnerabilities across versions and develop strategies to effectively address these issues. Understanding the root causes can lead to more secure coding practices.
\textbf{Enhancing Documentation Standards:} One notable observation from the resolved issue is its comprehensive accompanying information, which provides extensive CVE details about the vulnerabilities. The enhanced detail suggests that thorough documentation may play a pivotal role in the recognition and resolution of issues raised. Researchers might investigate how the quality and quantity of information impact the understanding and prioritization of vulnerabilities by developers. By setting clear, contextual, and actionable reporting standards, the research community can potentially increase the effectiveness of vulnerability management across the industry.

\subsection{Implications for Policymakers}


The findings indicate that static code analysis tools like CodeQL can serve as an objective measure of system security, making them valuable to regulatory bodies. Policymakers should consider:
\textbf{Incorporating Static Analysis in Standards:} Adopt static code analysis tools as a benchmark to evaluate the security of ADS and incorporate their use into industry standards and regulations.This can thereby protect consumers and promote public trust in autonomous driving technology.
\textbf{Promoting Security Best Practices:} Develop policies that encourage or require the use of static code analysis in the development lifecycle of ADS to foster a culture of security within the industry.
\textbf{Educating Stakeholders:} Provide guidelines and resources to help developers and organizations understand the benefits and proper use of static code analysis tools like CodeQL.

In conclusion, the implications of this research are significant, influencing the way developers approach code security, guiding researchers in their quest for more effective vulnerability detection methods, and informing policymakers in their efforts to regulate the industry. By embracing the findings of this study, all stakeholders can contribute to the advancement of ADS, ensuring that these systems are secure, reliable, and ready for widespread adoption.

\section{Threats to Validity} \label{Threats_to_validity}

In this section, we discuss the threats to the study's validity following the guidelines proposed by~\cite{runeson2009guidelines}, and how these threats were partially mitigated.

\textbf{Internal Validity.} One of the primary concerns in ensuring the internal validity of our study is to create a consistent environment for building each project successfully. Although we configured a clean and separate environment for each project, variations in requirements - such as different system versions and software dependencies - can affect the building process. Even when building different versions of the same project, this variation might lead to inconsistencies in the execution and output of CodeQL, which could in turn impact the accuracy of vulnerability detection. For example, if a third-party library is updated after a release of the project, the updates (such as stricter check mechanisms, modified functions and newly added patches) automatically installed when building the release might interfere the activity of vulnerability pinpointing. This can happen because the version constraints are not strictly defined, leading to unexpected behaviors. 
To mitigate this threat, we plan to conduct a larger-scale experiment that includes a broader range of environments and configurations and control as many environmental factors as possible. By doing so, we aim to capture a more comprehensive picture of how different build environments might affect the detection of vulnerabilities.


\textbf{External Validity.} It is important to acknowledge the potential biases come from our focus on three specific open-source ADS projects. This selection may not fully represent the broader landscape of proprietary, internally developed ADS that dominate the market, potentially limiting the generalization of our findings. 
The limited number of prominent open-source ADS projects is primarily due to the nascent stage of the autonomous driving industry, where many ADS projects are still under proprietary development by major automotive manufacturers and tech companies. These proprietary projects often have restricted access, making it challenging to conduct an in-depth analysis of their codebases. As a result, we focused on the most representative and accessible open-source projects that have gained significant attention and contributions from the community.
Moreover, the development of ADS projects evolves rapidly with continuous updates and patches, introducing a temporal dimension that could affect the timeliness of our results. The chosen time frame for observation is crucial to capture the current state of security vulnerabilities. While our study provides valuable insights into the security weaknesses of the selected projects, future research should consider expanding the scope to include a broader range of ADS projects as they become more openly available. This would help to further validate and generalize our findings across different development environments and project scales.
In summary, the limited number of target ADS projects in our study is a reflection of the current state of the ADS industry, where open-source projects with significant influence are still relatively rare. We believe that our findings provide a solid foundation for understanding the security vulnerabilities in prominent open-source ADS projects, and we encourage future research to build upon our work by exploring additional projects as they emerge in the open-source community.

\textbf{Construct Validity.} The query categories we have selected are based on industry standards and previous research, but these queries represent only a subset of the potential vulnerabilities that may exist in the code. This limitation could lead to an overestimation of the security of the ADS projects. To address this threat, employing a more systematic approach for query selection is needed.

\textbf{Conclusion Validity.} When tracking the life cycle of vulnerabilities, the dynamic nature of open-source projects presents many challenges. Updates on the projects can result in the relocation of code, potentially causing the difficult of tracking certain vulnerabilities. 
To confront this problem, we plan to enhance the tracking of vulnerability life cycles by integrating our detection results with the project commit histories. This measure will enable us to leverage commit messages and metadata to trace the evolution of code segments, thereby providing a more accurate and continuous monitoring of vulnerabilities as they are introduced, modified, or potentially resolved.

\section{Related Work} \label{related_work}
\subsection{Security of ADS}
The significance of ensuring the security of ADS cannot be overstated, as it directly impacts public safety, industry development, and societal acceptance of the technology. \cite{mariani2018overview} highlights the critical role of safety research in autonomous driving, acknowledging the presence of safety and security issues that regulatory bodies must address. The operational safety of AVs in real traffic environments and the types of security-related vulnerabilities are also emphasized, underscoring the pivotal nature of safety in public trust. Furthermore, advances in cloud services and big data analytics for enhancing traffic efficiency and safety, as mentioned by~\cite{ali2022big}, can be instrumental in bolstering the performance and security of ADS. Currently, data privacy protection and advanced data security technologies are gaining attention, especially in the context of vehicle communication networks~\citep{weimerskirch2010data}. Moreover, research on enhancing system robustness, such as through the application of Named Data Networking (NDN) in Vehicular Ad Hoc Networks (VANETs), is proved as essential for maintaining stability and resilience against dynamic changes and attacks~\citep{khelifi2019named}. The integration of Internet of Things (IoT) in intelligent transportation systems, as explored by~\cite{kaiwartya2016internet}, further emphasizing the need for robust data handling and analysis to ensure reliability and security.

However, various security threats pose challenges to the integrity and functionality of vehicles. These include AI safety issues 
being exploited by malicious actors, physical attacks on sensors, software vulnerabilities leading to system failures, and data pollution that can mislead decision-making processes~\citep{liu2020computing}. 
And ~\cite{den2018security} especially highlighted the importance of 
preventing adversarial attacks based on data pollution and potential software flaws. 
~\cite{lai2020security} further expanded on the security challenges in vehicular networks, including data privacy and protection against a spectrum of attacks. Meanwhile, a comprehensive study on software defects in AVs was conducted, analyzing 499 defects from two leading open-source ADS--Apollo and Autoware--and providing a classification system~\citep{garcia2020comprehensive}. However, similarly to many other studies, these researches focused on commits rather than directly examining the codebase. It was limited to vulnerabilities that had already been discovered and resolved, without offering warnings for potential, undiscovered vulnerabilities, and required a significant expenditure of human labor for analysis. Our research leverages CodeQL to systematically analyze vulnerabilities in ADS (examples can be found in Section \ref{examples}), contributing to a more detailed and comprehensive vulnerability distribution.

\subsection{Life Cycle and Duration of Vulnerabilities}
Researches have delved into the dynamics of how vulnerabilities are introduced, their causes, distribution, duration, and the typical methods employed for their detection and remediation. A significant finding from a large-scale analysis on open-source JavaScript projects revealed that vulnerabilities are often introduced during maintenance activities, such as bug fixes, and can persist in the codebase for an extended period, with an average lifetime of 511 days, increasing the window of opportunity for potential exploitation~\citep{bandara2021large}. And the study further underscores the criticality of the remediation phase, noting that in 90\% of the projects analyzed, commits fixing existing vulnerabilities inadvertently introduced new ones. These statistics underscore the complexity of the remediation process and the potential for new security issues to arise from attempted fixes. 

And the vulnerabilities that persist in the software can have significant implications beyond the immediate security risks. Undetected vulnerabilities lead to data breaches and system compromises, and in turn can erode trust among users~\citep{alomar2020you}.~\cite{frei2006large} highlighted the concerns of industry security practitioners about the lack of success in fixing vulnerabilities. 
This emphasizes the need for a robust vulnerability management process that includes not only the technical aspects of detection and remediation but also the organizational and procedural elements that support timely and effective responses. Our research builds upon this understanding, applying CodeQL to analyze the persistence and remediation of vulnerabilities in ADS, aiming to get a more targeted result of their life cycle, and thus enhance the efficacy of vulnerability management practices.

\subsection{Methods for Testing and Validating ADS}

Various testing methodologies have been employed to ensure the reliability and safety of complex technologies. Simulation testing, as discussed in many studies~\citep{li2020av, ramanagopal2018failing, wang2021advsim, fremont2020formal}, is crucial for identifying potential vulnerabilities and system failures in a risk-free setting. Automated vulnerability report analysis and genetic algorithms like AV-FUZZER have proven effective in simulating various traffic behaviors and identifying vulnerabilities~\citep{feng2019understanding, li2020av}. Meanwhile, real-world testing entails deploying actual vehicles or devices in operational settings to collect data and analyze behaviors under naturalistic conditions~\citep{pereira2019test}. The collected data also contributes to the foundation and baseline for simulation testing~\citep{karunakaran2023generating, fremont2020formal}. These methods help to understand the prevalence of IoT device vulnerabilities and their exploitation in real attacks, but cannot be performed very often because of their costs and dependencies on ADS as a whole.

The adoption of automated testing methods is further exemplified by~\cite{ferrara2021static}, who used static analysis to increase testing efficiency and reduce the reliance on expert human analysis as an approach to detect vulnerabilities in IoT devices. Static code analysis has also been recognized for other advantages, such as the ability to examine code systematically, allowing the identification of vulnerabilities that could potentially be missed during runtime testing~\citep{lyons2019towards}. The method faces challenges as being resource-intensive, and its complexity may lead to a high rate of false positives, which can undermine the confidence of developers in the efficacy of the tools~\citep{johnson2013don, ruthruff2008predicting}. This is partly because Rice's Theorem tells us that there are inherent limits to predicting program behavior perfectly. Due to the fact that non-trivial semantic properties of programs are undecidable, there is no algorithm that can determine them for all possible programs~\citep{xu2023review}. However, despite these limitations, attention is increasingly being paid in scenarios where code integrity and security are paramount, such as in the development of IoT systems, in which vulnerabilities can have severe consequences~\citep{abosata2021internet}. CodeQL, by monitoring the compilation process to perform static scans, possesses a degree of dynamism that mitigates some of the traditional issues associated with static analysis, offering a more reliable understanding of code behavior.

The empirical study conducted by~\cite{ayala2023empirical} reveals that among the top repositories on GitHub, only a fraction - 37\% for workflows and 7\% for security policies - is actively employing these critical security measures, indicating a significant room for improvement in the adoption of static analysis tools. The study also found that only 13.5\% of the top repositories that support CodeQL had it enabled, while the outcomes turned out to be beneficial~\citep{fischer2023effectiveness}. In Table \ref{tool_comparison}, while there are many static analysis tools like Snyk, Flawfinder, and SonarQube, they are limited by their language support, customization degree, and detection cost. Their scanning coverage does not fully satisfy our research requirements~\citep{snyk_docs, flawfinder_home, sonarqube_server_docs, codeql_cwe_coverage_docs, sonar_rules_database}. And the experience of developers on some of these tools (specifically, Snyk and SonarQube) also shows a significant decrease in accuracy compared to CodeQL
~\citep{lenarduzzi2020sonarqube, wu2023ossfp}. CodeQL, developed by GitHub, stands out for its ability to perform customized deep semantic analysis in the context of multilingual software systems~\citep{codeql_docs, youn2023declarative}, underscoring its practical value. Therefore, its effectiveness was evaluated and further utilized in our study to help bridge the gap between low cost and high quality of the testing approach.

\begin{table*}[]
\caption{Comparison of Different Static Analysis Tools for Security}
\label{tool_comparison}
\scalebox{1}{%
\begin{tabular}{llllllll}
\hline
\multirow{2}{*}{TOOLS} &
  \multirow{2}{*}{\begin{tabular}[c]{@{}l@{}}LANGUAGE\\ SUPPORT\end{tabular}} &
  \multirow{2}{*}{\begin{tabular}[c]{@{}l@{}}CUSTOMIZATION\\ DEGREE\end{tabular}} &
  \multirow{2}{*}{\begin{tabular}[c]{@{}l@{}}DETECTION\\ COST\end{tabular}} &
  \multirow{2}{*}{\begin{tabular}[c]{@{}l@{}}FULL CWE\\ COVERAGE\end{tabular}} &
  \multicolumn{3}{l}{\begin{tabular}[c]{@{}l@{}}SCANNING\\ COVERAGE\end{tabular}} \\ \cline{6-8} 
                &                                    &               &                 &              & cpp & py & js \\ \hline
Snyk            & C/C++, JavaScript, Python          & Medium        & Medium          & 106          & 14  & 15 & 14 \\ \hline
Flawfinder      & C/C++                              & Low           & Low             & 20           & 9   & 0  & 0  \\ \hline
SonarQube       & C/C++, JavaScript, Python          & High          & High            & 178          & 13  & 17 & 15 \\ \hline
\textbf{CodeQL} & \textbf{C/C++, JavaScript, Python} & \textbf{High} & \textbf{Medium} & \textbf{323} & \textbf{27} & \textbf{18} & \textbf{23} \\ \hline
\multicolumn{8}{p{450pt}}{Note. The `LANGUAGE SUPPORT' does not represent languages the tool able to scan, but lists the languages predominantly used in ADS that supported by it. The `SCANNING COVERAGE' means how many CWE categories can be detected through these tools in our planned testing list.
} \\
\end{tabular}%
}
\end{table*}

\subsection{Conclusive Summary}

The security of ADS is foundational to public trust and system integrity, yet it faces persistent challenges from AI exploits, sensor attacks, software flaws, and data pollution. Existing research, while highlighting critical vulnerabilities and remediation strategies, often lacks comprehensive codebase analysis or overlooks undiscovered risks. Vulnerabilities in ADS often emerge during maintenance and persist for extended periods
, underscoring the complexity of remediation and the risk of introducing new flaws. This necessitates robust vulnerability management that integrates technical, organizational, and procedural measures.

Testing methodologies, including simulation, real-world trials, and static analysis, are vital for identifying vulnerabilities. While simulation and real-world testing provide more direct links between security and ADS, their cost and static limitations hinder scalability. Static analysis tools have offered a more cost-effective choice, but they face limitations in language support and accuracy. CodeQL emerges as a superior solution, enabling deep semantic analysis across multilingual systems, bridging the gap between cost efficiency and high-quality testing. Its adoption, however, remains limited, signaling untapped potential in preemptive threat modeling. Our study leverages CodeQL to address these gaps, offering a granular view of vulnerability distribution and lifecycle dynamics. We advance the field of autonomous driving by systematically analyzing vulnerabilities, reinforcing the need for proactive, integrated security practices to enhance ADS resilience and reliability.

\section{Conclusions and Future Work} \label{Conclusions}

This study has provided a comprehensive analysis of the security vulnerabilities within ADS, revealing several key findings and contributions that have significant implications for the field. Our primary finding of prevalent CWE categories, such as CWE-190 (Integer Overflow or Wraparound) and CWE-20 (Improper Input Validation), underscores the need for targeted remediation efforts and a reevaluation of coding practices to address these recurring vulnerabilities. This insight is crucial for enhancing the security and reliability of ADS, as it identifies specific areas that require immediate attention and improvement.

Furthermore, our research highlights the persistence of vulnerabilities through version iterations, often remaining unnoticed or unaddressed for extended periods. This finding underscores the importance of continuous security auditing and the need for developers to maintain a proactive stance on vulnerability management. By integrating static code analysis tools like CodeQL into the development process, developers can automate the detection of potential vulnerabilities, thereby enhancing the quality of the code and reducing the risk of security breaches.

The empirical assessment of the impact of vulnerabilities on ADS performance offers a direct link between the severity of vulnerabilities and their tangible effects on system functionality. This finding reinforces the importance of timely remediation and the adoption of proactive security measures to ensure the safety and trustworthiness of ADS. The active response from the development community to the issues reported in our study further validates the effectiveness of CodeQL in uncovering actionable security issues, demonstrating the practical utility of static code analysis in the context of open-source ADS development.

Looking ahead, this research opens up new avenues for future studies. Potential directions include exploring the effectiveness of different static code analysis tools and developing hybrid approaches that combine the strengths of multiple tools to improve detection accuracy. Another area of interest could be investigating how the information in the static code analysis report affects the adoption and remediation of vulnerabilities in the development teams. Understanding these factors can lead to the creation of more effective communication mechanisms and the establishment of a security-conscious culture within the industry.

In conclusion, this work has not only advanced the understanding of security vulnerabilities in ADS but also provided actionable recommendations for improving the security and reliability of these systems. By integrating empirical research with practical strategies, this study aims to bolster the security of ADS and contribute to the broader goal of ensuring public safety and trust in autonomous driving technology.

\section*{Data availability}
We have shared the link to our dataset in the reference~\citep{CodeqlRes}.

\section*{Acknowledgments}
This work was funded by the National Natural Science Foundation of China under Grant Nos. 62176099 and 62172311, the Major Science and Technology Project of Hubei Province under Grant No. 2024BAA008. The numerical calculations in this paper have been done on the supercomputing system in the Supercomputing Center of Wuhan University.

\printcredits

\bibliographystyle{cas-model2-names}
\bibliography{ref}

\balance

\end{sloppypar}
\end{document}